\documentclass{LMCS}

\def\dOi{10(3:13)2014}
\lmcsheading%
{\dOi}
{1--13}
{}
{}
{Oct.~\phantom07, 2013}
{Sep.~\phantom02, 2014}
{}

\ACMCCS{[{\bf Mathematics of computing}]: Mathematical analysis---Numerical analysis}


\usepackage{amssymb}
\usepackage{amsmath}
\usepackage{latexsym,hyperref}

\def \dom{{\rm dom}}

\def \In{{\subseteq}}
\def \om{{\Sigma^\omega }}
\def \pf{:\hspace{0.6ex}\subseteq \hspace{-0.1ex}}
\def \s{{\Sigma^*}}
\def \dd {{\:\rm d}}
\def \one{{1\hspace{-.65ex}{\rm I}}}

\def\IB{{\mathbb{B}}}
\def\IN{{\mathbb{N}}}
\def\IQ{{\mathbb{Q}}}
\def\IR{{\mathbb{R}}}

\newcommand{\an}{\ \ \mbox{and}\ \ }

\newcommand{\pproof}{\noindent{\it Proof. } }

\newcommand{\bb}{ \hspace{-0.76ex}- \hspace{-0.80ex}}

\def \BV {{\rm BV}}
\def \BM {{\rm BM}}
\def \CF {{\rm CF}}
\def \Cp {{C'[0;1]}}

\begin{document}

\title[Computable Jordan decomposition]{Computable Jordan Decomposition of Linear Continuous Functionals on {\boldmath $C[0;1]$}}

\author[T.~Jafarikhah]{Tahereh Jafarikhah\rsuper a}   
\address{{\lsuper a}University of Tarbiat Modares, Tehran, Iran} 
\email{t.jafarikhah@modares.ac.ir}  

\author[K.~Weihrauch]{Klaus Weihrauch\rsuper b} 
\address{{\lsuper b}Dpt. of Mathematics and Computer Science, University of Hagen, Germany}    
\email{Klaus.Weihrauch@FernUni-Hagen.de}  



\begin{abstract}

By the Riesz representation theorem using the Riemann-Stieltjes
integral, linear continuous functionals on the set of continuous
functions from the unit interval into the reals can either be
characterized by functions of bounded variation from the unit interval
into the reals, or by signed measures on the Borel-subsets. Each of
these objects has an (even minimal) Jordan decomposition into
non-negative or non-decreasing objects. Using the representation
approach to computable analysis, a computable version of the Riesz
representation theorem has been proved by Jafarikhah, Lu and
Weihrauch. In this article we extend this result. We study the
computable relation between three Banach spaces, the space of linear
continuous functionals with operator norm, the space of (normalized)
functions of bounded variation with total variation norm, and the
space of bounded signed Borel measures with variation norm. We
introduce natural representations for defining computability. We prove
that the canonical linear bijections between these spaces and their
inverses are computable.  We also prove that Jordan decomposition is
computable on each of these spaces.
\end{abstract}

\maketitle

\section{Introduction}\label{seca}
Let $C[0;1]$ be the set of continuous functions $h:[0;1]\to\IR$.
By the Riesz representation theorem for every linear continuous
function $F:C[0;1]\to\IR$ there is a function $g:[0;1]\to \IR$ of
bounded variation such that $F(h)=\int h\dd g$ for every
continuous function $h\in C[0;1]$. For every function  $g:[0;1]\to
\IR$ of bounded variation there is a signed Borel measure $\mu$ on
the unit interval of finite variation norm such that $\int h\dd g
=\int h\dd\mu $ for every continuous function $h\in C[0;1]$.
Finally for every signed Borel measure $\mu$ on the unit interval
of finite variation norm the function $h\mapsto \int h\dd\mu$ for
$h\in C[0;1]$ is linear and continuous.

By the Jordan decomposition theorem for every function $g:[0;1]\to
\IR$ of bounded variation there are non decreasing functions
 $g^+, g^-:[0;1]\to\IR$ such that $g=g^+-g^-$. Similar decomposition theorems have been proved for Functionals $F$ and measures $\mu$: for
 every linear continuous functional $F:C[0;1]\to\IR$ there are two non-negative functionals $F^+$ and $F^-$ such that $F=F^+-F^-$, and
for every signed Borel measure $\mu$ on the unit interval of
finite variation norm there are non-negative measures
$\mu^+,\mu^-$ such that $\mu=\mu^+-\mu^-$. In each case there is a
minimal decomposition
\cite{Doo94,Dud02,GP65,KK96,KF75,Sch97b,Coh80,Bog07-1,Bog07-2,Nat75}.

In this article we study computability of all of these existence
theorems. Computability of the Riesz representation theorem and
its converse have been proved in \cite{LW07a} with a revised proof
in \cite{JW13}. Computability of $(\mu,h)\mapsto \int h\dd \mu$
for continuous $h$ and non-negative bounded Borel measure $\mu$
has been proved in \cite{Wei99a}. In this article we extend these
results.

We study the computable relation between three Banach spaces, the
space of linear continuous functionals with operator norm, the
space of (normalized) functions of bounded variation with total
variation norm, and the space of bounded signed Borel measures
with variation norm. We introduce natural representations for
defining computability. We prove that the canonical linear
bijections $F\mapsto g$, $g\mapsto\mu$ and $\mu\mapsto F$ between
these spaces and their inverses are computable. We also prove that
(minimal) Jordan decomposition is computable on each of these
spaces.

In Section~\ref{secq} we summarize some definitions and basic
facts from classical analysis on linear continuous functionals
$F:C[0;1]\to \IR$, functions of bounded variation and the
Riemann-Stieltjes integral, and on signed measures on the Borel
sets of the unit interval. We consider only functions
$g:[0;1]\to\IR$ of bounded variation which are normalized in the
sense that $g(0)=0$ and for all $0<y<1$, $\lim_{x\nearrow
y}g(x)=g(y)$.

In Section~\ref{secr} we outline very shortly some general
concepts from the representation approach to computable analysis
\cite{Wei00,BHW08}. For defining computability we introduce and
discuss representations of the functionals, of the functions of
bounded variation and of the signed measures  and also
representations of the subspaces of non-negative or non-decreasing
objects, respectively. While in \cite{LW07a,JW13} partial
functions of bounded variation are considered in this article we
use total normalized functions with a representation which is very
closely related to the one used for the partial functions.

In Section~\ref{secs} first we prove for the special case of
non-negative functionals $F$, non-decreasing functions $g$ and
non-negative measures $\mu$ that the mappings $F\mapsto g$,
$g\mapsto\mu$ and $\mu\mapsto F$ such that $F(h)=\int h\dd g$,
$\int h\dd g=\int h\dd\mu$ and $\int h\dd\mu=F(h)$ are computable
w.r.t the ``non-negative'' representations. Then we prove our main
results: On the spaces of linear continuous functionals with
operator norm, the space of normalized functions of bounded
variation with variation norm and the space of signed measures
with finite variation norm the operators $F\mapsto g$,
$g\mapsto\mu$ and $\mu\mapsto F$ are computable. Furthermore,
 the Jordan decompositions $F\mapsto (F^+,F^-)$, $ g\mapsto (g^+,g^-)$ and $\mu\mapsto(\mu^+,\mu^-)$ are computable.
The results can be expressed in such a way that a number of
representations of the space of linear continuous functionals are
equivalent.

The results can be generalized easily from the unit interval to
arbitrary intervals $[a;b]$ with computable endpoints. More
generally, the results can be proved computably uniform in $a,b$,
where $a$ and $b$ are given by their standard representation via
fast converging Cauchy sequences of rational numbers.

In \cite{Ko91,ZR05a} Jordan decomposition of computable real
functions  and of polynomial time computable functions  on the
unit interval has been studied. However, they do not investigate
computability of the Jordan decomposition operator but ask whether
computability or polynomial computability is preserved under
Jordan decomposition. Ko \cite{Ko91} has shown that there is a
polynomial time computable function $f$ of bounded variation which
is not the difference of two non-decreasing polynomial time
computable functions. This has been strengthened by Zheng and
Rettinger who have proved  that there is a polynomial time
computable function of bounded variation with polynomial modulus
of absolute continuity which is not the difference of two
non-decreasing computable functions.

\section{Basics from the classical theory}\label{secq}
We summarize some definitions and results about functions of
bounded variation and from (non-computable) measure theory which
are scattered across many sources
\cite{Doo94,Dud02,GP65,KK96,KF75,Sch97b,Coh80,Bog07-1,Bog07-2,Nat75,LW07a,JW13}
or can be derived easily from there. For convenience we consider
only the closed unit interval $[0;1]$ for functions,  measures
etc.

Let $C[0;1]$ be the space of continuous functions $h:[0;1]\to\IR$
with norm $\|h\|=\sup\{|h(x)|\mid x\in[0;1]\}$. Let $C'[0;1]$ be
the space of linear continuous functionals $F:C[0;1]\to\IR$ with
norm $\|F\|=\sup\{ |F(h)|\mid h\in C[0;1], \|h\|\leq 1\}$. For
every non-negative $F\in C'[0;1]$ (that is, $F(h)\geq 0$ if $h\geq
0$), $\|F\|=F(\one)$ (where $\one(x)=1$ for $0\leq x\leq 1$).

We shortly introduce functions $g:[0,1]\to\IR$ of bounded
variation and the Riemann-Stieltjes integral $\int h\dd g$ for
continuous functions $h:[0;1]\to\IR$. A {\em partition} of a real
interval $[a;b]$ ($a<b$) is a sequence $Z=(x_0,x_1,\ldots,x_n)$,
$n\geq 0$, of real numbers such that $a=x_0<x_1\ldots <x_n=b$. The
partition $Z$ has {\em precision} $k$, if $x_i-x_{i-1}< 2^{-k}$
for $1\leq i\leq n$. A partition $Z'=(x'_0,x'_1,\ldots,x'_m)$, is
finer than $Z$, if
$\{x_0,x_1,\ldots,x_n\}\In\{x'_0,x'_1,\ldots,x'_m\}$. For a
function $g:[0;1]\to\IR$, for $0\leq a<b\leq 1$ and a partition
$Z$ of the interval $[a;b]$ define
\begin{eqnarray}\label{f17}
S(g,Z) &:= & \sum_{i=1}^n|g(x_i)-g(x_{i-1})|\,,\\
\label{f18}
V_a^b(g) & := & \sup \{S(g,Z)| Z\ \mbox{is a partition of } [a;b]\}\,.
\end{eqnarray}
The function $g:[0;1]\to\IR$ is of {\em bounded variation} if its
variation ${\rm Var}(g):=V_0^1(g)$ is finite. For a function of
bounded variation the total variation function $/g/:[0;1]\to \IR$
is defined by $/g/(0):=0$ and $/g/(x):=V_0^x(g)$.

In the following let $h:[0;1]\to \IR$ be a continuous function and
let $g:[0;1]\to\IR$ be a function of bounded variation.
 For any partition
$Z=(x_0,x_1,\ldots,x_n)$ of $[0;1]$ define
%
\begin{eqnarray}\label{f25}
S(g,h,Z):= \sum_{i=1}^nh(x_i)(g(x_i)-g(x_{i-1})).
\end{eqnarray}
%
 Since $h$ is continuous and its domain is compact, it has a (uniform) {\em modulus of continuity}, i.e.,
a function $m:\IN\to\IN$ such that  $|h(x)-h(y)|\leq 2^{-k}$ if
$|x-y|\leq 2^{-m(k)}$. We may assume that the function $m$ is
non-decreasing.

\begin{lem}[\cite{LW07a}]\label{l5}
Let $h:[0;1]\to \IR$ be a continuous function with modulus of
continuity $m:\IN\to\IN$ and let $g$ be a function of bounded
variation.  Then there is a unique number $I\in \IR$ such that
\[|I-S(g,h,Z)|\leq 2^{-k}{\rm Var}(g)
\]
 for all $k\in\IN$ and for every partition $Z$ of $[0;1]$ with precision $m(k+1)$.
\end{lem}
%
 The number $I$ from
Lemma~\ref{l5} is called the {\em Riemann-Stieltjes integral} and
is denoted by $\int h\, {\rm d}g$. The operator $F_g:h\mapsto \int
h\dd g$ is linear and continuous on $C[0;1]$.

Notice that by Lemma \ref{l5} the integral $\int h\dd g$ is
determined already by the values of the function $g$ on $0$ and
$1$ and on an arbitrary dense set $X$, since there are partitions
of arbitrary precision that contain points only from the set~$X$.
If $g$ is of bounded variation, then $\lim_{y\nearrow x}g(y)$ and
$\lim_{y\searrow x}g(y)$ exist for all $0\leq x\leq 1$. Functions
of bounded variation can be normalized without changing the
Riemann-Stieltjes integral over  continuous functions.
\medskip

Let ${\rm BV}$ be the set of functions $g:[0;1]\to\IR$ of bounded
variation such that
\begin{eqnarray}
g(0)=0 \mbox{ and }
(\forall \  0<x<1)\;g(x)=\lim_{y\nearrow x}g(y)\,.
\end{eqnarray}

\begin{lem}\label{l8}$ $
\begin{enumerate}
\item\label{l8c} Every $g\in{\rm BV}$ is left-continuous.
\item\label{l8a} For every $g\in{\rm BV}$, ${\rm Var}(g)=\|
    F_g\|$.
\item\label{l8b} For every function $g$ of bounded variation
    there is a unique function $g'\in{\rm BV}$ such that $\int
    h \dd g= \int h \dd g'$ for all functions $h\in C[0;1]$.
\end{enumerate}
\end{lem}

\noindent The function $g'$ can be defined by
\begin{eqnarray}\label{f28}\mbox{$g'(0):=0$, $g'(1):=g(1)-g(0)$ and  $g'(x):=\lim_{y\nearrow x}g(y)-g(0)$ for $0<x<1$.}
\end{eqnarray}
For every non-decreasing function $g\in\BV$, ${\rm Var}(g)=g(1)$.
\medskip

Let   $\BM$ be the set of signed measures $\mu$ with finite
variation norm $\|\mu\|_m$ on the Borel subsets of the unit
interval $[0;1]$ defined by $\|\mu\|_m:=\sup_\pi \sum_{I\in\pi}
|\mu(I)|$ where $\pi$ runs over all finite partitions of the unit
interval into intervals (open, semi-open, closed). If $\mu$ is
non-negative, then $\|\mu\|_m=\mu([0;1])$.

The following theorem summarizes the relation between the three
spaces introduced above.

\begin{thm} \label{t2}  The spaces $(\Cp,\|\,.\,\|)$, $(\BV,{\rm Var})$ and $(\BM,\|\,.\,\|_m)$ are Banach spaces.
\begin{enumerate}

\item \label{t2a} There is a unique linear homeomorphism
    $T_{\rm FV}:\Cp\to\BV$ such that \\$T_{\rm FV}(F)=g$
    implies \ $(\forall h\in C[0;1])\,F(h)=\int h\dd g$.

\item \label{t2b} There is a unique linear homeomorphism
    $T_{\rm VM}:\BV\to\BM$ such that \\$T_{\rm VM}(g)=\mu$
    implies $(\forall h\in C[0;1])\int h\dd g=\int h\dd\mu$.

\item \label{t2c} There is a unique linear homeomorphism $T_{\rm MF}:\BM\to\Cp$ such that \\
    $T_{\rm MF}(\mu)=F$ implies $(\forall h\in C[0;1])\int
    h\dd\mu =F(h)$.
\end{enumerate}
The functions $T_{\rm FV}$,  $T_{\rm VM}$  and   $T_{\rm FV}$
preserve the norms. Moreover, if $F$  is non-negative then $T_{\rm
FV}(F)$ is non-decreasing, if $g$  is non-decreasing then $T_{\rm
VM}(g)$ is non-negative, and if $\mu$ is non-negative then $T_{\rm
MF}(\mu)$ in non-negative.
\end{thm}

The three spaces are not separable. Theorem~\ref{t2}(\ref{t2a}) includes the Riesz representation
theorem \cite{GP65}. For  real numbers $x$ define $x^+:=(|x|+x)/2$
and $x^-:=(|x|-x)/2$. Then $x^+$ and $x^-$ are non-negative
numbers such that $x=x^+-x^-$. Moreover, $x^+$ and $x^-$ are
minimal, that is,  $x^+\leq y^+$ and $x^-\leq y^-$ if $y^+,y^-$
are non-negative such that $x=y^+-y^-$. By the Jordan
decomposition theorem, this kind of decomposition can be
generalized to functionals $F\in C'[0;1]$, to functions $g\in \BV$
and to signed measures $\mu\in \BM$.

\begin{defi}\label{d10}$ $
\begin{enumerate}
\item For $F\in \Cp$ the {\em Jordan decomposition} is a pair
    $(F^+,F^-)$ of non-negative functionals in $C'[0;1]$ such
    that $F=F^+-F^-$, and if $\,G^+,G^-\in C'[0;1]$ are
    non-negative functionals such that $F=G^+ -G^-$ then
    $F^+\leq G^+$ and $F^-\leq G^-$.
\item For $g \in \BV$ the {\em Jordan decomposition} is a pair
    $(g ^+, g^-)$ of non-decreasing functions in $\BV$ such
    that $g=g^+-g^-$, and if \;$t^+, t^-\in\BV$ are
    non-decreasing functions such that $\:g=t^+ -t^-$ then
    $g^+\leq t^+$ and $g^-\leq t^-$.

\item For $\mu\in \BM$ the {\em Jordan decomposition} is a
    pair $(\mu^+,\mu^-)$ of non-negative measures in $\BM$
    such that $\mu=\mu^+ -\mu^-$, and if $\;\nu^+,\nu^-\in
    \BM$ are non-negative measures such that $\mu=\nu^+
    -\nu^-$ then  $\mu^+\leq\nu^+$ and $\mu^-\leq\nu^-$.
\end{enumerate}
\end{defi}

\noindent If a Jordan decomposition exists then it is unique by the
minimality condition. Notice that some authors do not require
minimality  for Jordan decomposition.

\begin{thm}\label{t14}$ $
\begin{enumerate}

\item Every $F\in C'[0;1]$ has a Jordan decomposition. If $F^+,F^-\in\Cp$ are non-negative and $F=F^+-F^-$, then \\
    $(F^+,F^-)$ is the Jordan decomposition of $F$ iff
    $\,\|F\|=\|F^+\| +\| F^-\|$.

\item Every $g \in\BV$ has a Jordan decomposition. If $g^+,g^-\in\BV$ are non-decreasing and $g=g^+-g^-$, then\\
    $(g^+,g^-)$ is the Jordan decomposition of $g$ iff $\,{\rm
Var}(g)={\rm Var}(g^+) + {\rm Var}(g^-)$.
\item Every measure $\mu\in \BM$ has a Jordan decomposition.
    If
$\mu^+, \mu^-\in\BM$ are non-negative measures and $\mu=\mu^+ -\mu^-$, then \\
    $(\mu^+,\mu^-)$ is the Jordan decomposition of $\mu$ iff
    $\,\|\mu\|_m=\|\mu^+\|_m + \|\mu^-\|_m$.
\end{enumerate}
\end{thm}

\begin{cor}\label{c3}$ $
\begin{enumerate}
\item  \label{c3a} If $(F^+,F^-)$ is the Jordan decomposition
    of $F$ then $(T_{\rm FV}(F^+),(T_{\rm FV}(F^-))$ is the
    Jordan decomposition of  $T_{\rm FV}(F)$.
\item  \label{c3b} If $(g^+,g^-)$ is the Jordan decomposition
    of $g$ then $(T_{\rm VM}(g^+),(T_{\rm VM}(g^-))$ is the
    Jordan decomposition of  $T_{\rm VM}(g)$.
\item  \label{c3c} If $(\mu^+,\mu^-)$ is the Jordan
    decomposition of $\mu$ then $(T_{\rm MF}(\mu^+),(T_{\rm
    MF}(\mu^-))$ is the Jordan decomposition of  $T_{\rm
    MF}(\mu)$.
\end{enumerate}
\end{cor}

\pproof Let $(F^+,F^-)$ be the Jordan decomposition of
$F:=F^+-F^-$. Let $g^+:=T_{\rm FV}(F^+)$ and $g^-:=T_{\rm
FV}(F^-)$. Then $g^+-g^-= T_{\rm FV}(F^+-F^-)$. By
Theorems~\ref{t2} and~\ref{t14},

${\rm Var}(g^+-g^-)=\|F^+-F^-\| =\|F^+\| +\|F^-\|= {\rm Var}(g^+)
+
{\rm Var}(g^-)$,\\
hence by Theorem~\ref{t14}, $(g^+,-g^-)$ is the Jordan
decomposition of $(g^+-g^-)$. Therefore, $(T_{\rm FV}(F^+),T_{\rm
FV}(F^-))$ is the Jordan decomposition of $T_{\rm FV}(F)$.

The other statements can be proved accordingly. \qed

\section{The concepts of computability}\label{secr}
In this section we define computability on the three spaces from
Theorem~\ref{t2}. Since the spaces are not separable, Cauchy
representations \cite[Chapter~8.1]{Wei00} are not available.

For studying computability we use the representation approach
(TTE, Type 2 Theory of Effectivity) for computable analysis
\cite{Wei00,BHW08}. Let $\Sigma$ be a finite alphabet. Computable
functions on $\s$ (the set of finite sequences over $\Sigma$) and
$\om$ (the set of infinite sequences over $\Sigma$)  are defined
by Turing machines which map sequences to  sequences (finite or
infinite). On $\s$ and $\om$ finite or countable tuplings
(injections from cartesian products of $\s$ and $\om$ to $\s$ or
$\om$) will be denoted by $\langle\ \rangle$
\cite[Definition~2.1.7]{Wei00}. The tupling functions and the
projections of their inverses are computable.

In TTE, sequences from $\s$ or $\om$ are used as ``names'' of
abstract objects such as rational numbers, real numbers, real
functions or points of a metric space. We consider computability
of multi-functions w.r.t. representations \cite{Wei00,BHW08},
\cite[Sections~3,6,8,9]{Wei08}. A representation of a set $X$ is a
function $\delta\pf C\to X$ where $C=\s$ or $C=\om$. If
$\delta(p)=x$ we call $p\,$ a $\delta$-{\em name} of $x$.

For representations $\gamma\pf Y \to M$ and $\gamma_0\pf Y_0\to
M_0$, a function $h\pf Y\to Y_0$ is a
$(\gamma,\gamma_0)$-realization of a function $f\pf M\to M_0$, iff
for all $p\in Y$ and $x\in M$,
\begin{eqnarray}\label{f3}
 \gamma(p)=x\in \dom(f) & \Longrightarrow & \gamma_0\circ h(p)= f(x)\,,
\end{eqnarray}
that is, $h(p)$ is a name of some $ f(x)$, if $p$ is a name of
$x\in\dom(f)$.
The function $f$ is called $(\gamma,\gamma_0)$\bb computable, if
it has a computable $(\gamma,\gamma_0)$-realization and
$(\gamma,\gamma_0)$-continuous if it has a continuous realization.
The definitions can be generalized straightforwardly to
multivariate functions $f\pf M_1\times \ldots\times M_n\to M_0$
for represented sets $M_i$.

For two representations $\delta_i\pf Y_i\to M_i$ ($i=1,2$),
 $\delta_1$ is {\em reducible to} $\delta_2$, $\delta_1\leq \delta_2$,  iff there is a computable function $h:\In Y_1\to Y_2$ such that
$(\forall\,p\in\dom(\delta_1))\,\delta_1(p) =\delta_2h(p)$ (if $p$
is a $\delta_1$-name of $x$ then $h(p)$ is a $\delta_2$-name of
$x$). The two representations are {\em equivalent},
$\delta_1\equiv\delta_2$, iff $\delta_1\leq\delta_2$ and
$\delta_2\leq\delta_1$.

 Let $\delta_i\pf \om \to M_i$ ($i=1,2$) be representations.
The canonical representation $[\delta_1,\delta_2]$ of the product
$M_1\times M_2$ is defined by
\begin{eqnarray}\label{f41}
[\delta_1,\delta_2]\langle p_1,p_2\rangle &= & (\delta_1(p_1),\delta(p_2))\,.
\end{eqnarray}
There is a representation $[\delta_1\to\delta_2]$ of the set of
$(\delta_1,\delta_2)$-continuous functions $f:M_1\to M_2$  which
is determined uniquely up to equivalence by $({\bf U})$ and $({\bf
S})$\cite{Wei00}.
\begin{eqnarray}\label{f42}
 \mbox{{\bf (U)} The apply function $(f,x)\mapsto f(x)$ is $([\delta_1\to\delta_2],\delta_1,\delta_2)$-computable,}
\end{eqnarray}
\vspace{-2ex}
\begin{eqnarray}\label{f43}
\begin{array}{l}
\mbox{{\bf (S)} If for some representation $\gamma$  of a set of $(\delta_1,\delta_2)$-continuous functions}\\
\mbox{ \ \ \   $\ (f,x)\mapsto f(x)$ is $(\gamma,\delta_1,\delta_2)$-computable
 then $\gamma\leq [\delta_1\to\delta_2]$.}
\end{array}
\end{eqnarray}

\noindent $({\bf U})$ corresponds to the ``universal Turing machine
theorem'' and $({\bf S})$ to the ``smn-theorem'' from
computability theory. Roughly speaking, $[\delta_1\to\delta_2]$ is
(up to equivalence) the ``weakest'' representation of the set of
$(\delta_1,\delta_2)$-continuous functions for which the apply
function is computable. The generalized Turing machines in
\cite{TW11b} are useful tools for defining new computable
functions on represented sets from given ones.

We use various canonical notations $\nu\pf \s\to X$: $\nu_\IN$ for
the natural numbers, $\nu_\IQ$ for the rational numbers, $\nu_{\rm
Pg}$ for the polygon functions on $[0;1]$ whose graphs have
rational vertices, and $\nu_I$ for the set  $\rm RI$ of open
intervals $(a;b)\In (0;1)$ with rational endpoints. For functions
$m:\IN\to\IN$ we use the canonical representation
$\delta_\IB\pf\om\to\IB=\{m\mid m:\IN\to\IN\}$ defined by
$\delta_\IB(p)=m$ if $p=1^{m(0)}01^{m(1)}01^{m(2)}0\ldots$. For
the real numbers we use the Cauchy representation $\rho\pf
\om\to\IR$, $\rho(p)=x$ if $p$ is (encodes) a sequence
$(a_i)_{i\in\IN}$ of rational numbers such that for all $i$,
$|x-a_i|\leq 2^{-i}$, and the lower representation $\rho_<$,
$\rho_<(p)=x$ iff $p$ is (encodes) a sequence $(a_i)_{i\in\IN}$ of
rational numbers such that $x=\sup_ia_i$. By the Weierstra{\ss}
approximation theorem the countable set $\rm Pg$ of polygon
functions with rational vertices is dense in $C[0;1]$. Therefore,
$C[0;1]$ with notation $\nu_{\rm Pg}$ of the set $\rm Pg$ is a
computable metric space \cite{Wei00} for which we use the Cauchy
representation $\delta_C$ defined as follows: $\delta_C(p)=h$ if
$p$ is (encodes) a sequence $(h_i)_{i\in\IN}$ of polygons
$h_i\in{\rm Pg}$ such that for all $i$, $\|h-h_i\|\leq 2^{-i}$
\cite{Wei00}.
\smallskip

Since the representations $\rho$ and $\delta_C$ are admissible, a
functional $G:C[0;1]\to \IR$ is continuous iff it is
$(\delta_C,\rho)$-continuous \cite{Wei00}. Therefore,
$[\delta_C\to\rho]$ is a representation of the continuous
functionals $G:C[0;1]\to \IR$. This representation  is tailored
for evaluation $(G,h)\mapsto G(h)$ (\ref{f42}) (\ref{f43}). We use
it for the subspace $\Cp$ of the linear continuous functionals.
The norm on $\Cp$ is $([\delta_C\to\rho],\rho_<)$-computable but
not $([\delta_C\to\rho],\rho)$-computable. Since for computations
we will need the $\rho$-name of the norm we include it in the
name.

\begin{defi}\label{d8} Define a representation $\delta_\CF$ of $\Cp$ by
\[\delta_\CF\langle p,q\rangle=F:\iff [\delta_C\to\rho](p)=F \an \rho(q)=\|F\|\,.\]
\end{defi}

This is the representation of the dual of $C[0;1]$  space as
suggested in Section 15 (see also Definition 3.9) of [V.Brattka:
"Computability of Banach Space Principles"] in the case that this
dual is not separable. It is admissible and admits computability
of scalar multiplication, the norm and the rapid Lim-operator, but
vector addition is not computable. This yields a good
justification for using $\delta_{\rm CF}$.

In \cite{LW07a,JW13} a computable version of the Riesz
representation theorem is proved. In these articles the concept of
bounded variation is generalized straightforwardly to the set
${\rm BVC}$ with representation $\delta_{BVC}$ of partial
functions $g\pf [0;1]\to \IR$ with countable dense domain
containing $\{0,1\}$ which are continuous on $\dom(g)\setminus
\{0,1\}$. Remember that a function of bounded variation has at
most countably many points of continuity. The integral $\int h\dd
g$ for continuous $h$ and an arbitrary function $g$ of bounded
variation is defined already by any restriction of $g$ to a
countable dense subset containing $\{0,1\}$ \cite{JW13}. Every
(partial) function $g\in {\rm BVC}$ can be extended uniquely to a
normalized (total) function ${\rm ext}(g)\in\BV$ by ${\rm
ext}(g)(x):=\lim_{y\nearrow x,\ y\in \dom(g)}g(y)$ for
$x\not\in\dom(g)$. Then $\int h\dd g =\int h\dd \ {\rm ext}(g)$
for all $h\in C[0;1]$ an ${\rm Var}(g)={\rm Var}({\rm ext}(g))$.
In this article instead of $\delta_{\rm BVC}$ we use the
representation $\delta_V:={\rm ext}\circ \delta_{\rm BVC}$ of the
normalized functions. The variation is not
$(\delta_V,\rho)$-computable but only
$(\delta_V,\rho_<)$-computable. Since for computations we will
need the $\rho$-name of the variation we include it in the name.
Notice that for computing the Riemann-Stieltjes integral $\int
h\dd g$  a $\delta_V$-name and an upper bound of ${\rm Var}(g)$
suffice \cite{LW07a,JW13}.

\begin{defi}\label{d7}
Define representations $\delta_{\rm V}$ and $\delta_\BV$ of
$\,\BV$ as follows:  \begin{enumerate}
\item $\delta_{\rm V}(p)=g$ iff there are $p_0,q_0,p_1,q_1,\ldots
    \in\om$ such that $p=\langle \langle p_0,q_0\rangle,\langle
    p_1,q_1\rangle,\ldots\rangle$, $\rho(p_0)=\rho(q_0)=0$,
    $\rho(p_1)=1$, $g\circ \rho(p_i)=\rho(q_i)$ for all $i\in\IN$,
    $A_p:=\{\rho(p_i)\mid i\geq 2\}$ is a dense subset of $(0;1)$
    and $g$ is continuous on $A_p$.

\item $\delta_\BV\langle p,q\rangle =g :\iff \delta_{\rm V}(p)=g
    \an \rho(q)={\rm Var}(g)$.
\end{enumerate}
\end{defi}

A computable version of the Riesz representation theorem and its
converse have been proved in \cite{LW07a,JW13}. The results can be
formulated  as follows.

\begin{thm}[Computable Riesz representation \cite{LW07a,JW13}]\label{t5}$ $\\
The function $(F,z)\mapsto g$  mapping every functional $F\in
C'[0;1]$ and its norm $z$ to the (unique)  function $g\in \BV$
such that $F(h)=\int h\dd g$ (for all $h\in C[0;1]$) is
$([\delta_C\to\rho],\rho,\delta_{\rm V})$-computable.
\end{thm}

\begin{thm}[\cite{LW07a,JW13}]\label{t8}
The operator $(g,l)\mapsto F$, mapping every $g\in{\rm BV}$ and
every $l\in\IN$ with  ${\rm Var}(g)\leq 2^l$ to the functional $F$
defined by $F(h)= \int h\dd g$ for all $h\in C[0;1]$, is
$(\delta_{\rm V},\nu_\IN,[\delta_C\to\rho])$-computable.
\end{thm}

By a slight generalization of the representation $\delta_m$ of the
probability measures on the Borel sets of the unit interval
defined and studied in \cite{Wei99a} we obtain a representation of
the bounded non-negative Borel measures on the unit interval. Let
${\rm Int}:=\{(a,b), [0;b), (a;1], [0;1]\mid a,b\in\IQ, 0\leq
a<b\leq 1\}$ be the set of all rational open subintervals of
$[0;1]$.

\pagebreak

\begin{defi}\label{d6} Let $\BM_+$ be the set of non-negative bounded measures.

\begin {enumerate}
\item Define a representation $\delta_m$ of the set ${\rm BM}_+$
    as follows. For $p,q\in\om$ and $\mu\in{\rm BM}_+$,
    $\delta_m\langle p,q\rangle=\mu$ iff $\rho(q)=\mu([0;1])$ and
    $p$ is (encodes) a list of all $(a,J)\in\IQ\times {\rm Int}$
    such that $a<\mu(J)$.
\item Define a representation of $\BM$ by $\delta_\BM\langle
    p,q,r\rangle=\mu$ iff $\mu=\delta_m(p)-\delta_m(q)$ and
    $\|\mu\|_m=\rho(r)$.
\end{enumerate}
\end{defi}
Roughly speaking by this definition, $\delta_m$ is the greatest
(or "poorest") representation $\gamma$ of the bounded non-negative
measures such that $\mu([0;1])$ can be computed and $a<\mu(J)$
($a\in\IQ$ and $J\in{\rm Int}$) can be enumerated. By the next
theorem the representation $\delta_m$  is the greatest
representation of the non-negative bounded measures for which
$\mu([0;1])$ and integration of continuous functions are
computable.

\begin{thm}\label{t7}$ $
\begin{enumerate}
\item \label{t7a}The function $\mu\mapsto \mu([0;1])$ is
    $(\delta_m,\rho)$-computable, and the function
    \\$(\mu,h)\mapsto \int h\dd\mu$ is
    $(\delta_m,\delta_C,\rho)$-computable.
\item \label{t7b}If for some representation $\gamma$ of ${\rm
    BM}$ the function $\mu\mapsto \mu([0;1])$ is
    $(\gamma,\rho)$-computable, and the function
    $(\mu,h)\mapsto \int h\dd\mu$ is
    $(\gamma,\delta_C,\rho)$-computable then
    $\gamma\leq\delta_m$.
\end{enumerate}
\end{thm}

\pproof
\begin{enumerate}
 \item The first statement is obvious, the second one can be derived easily from the special case for measures with $\mu([0;1])=1$ \cite[Theorem~3.6]{Wei99a}.

 \item This can be deduced from \cite[Theorem~4.2]{Wei99a}.\qed
\end{enumerate}

 \section{Computable equivalence of the three concepts and computable Jordan decomposition}\label{secs}
We will now apply the representations introduced in
Section~\ref{secr}:
\\
-- \ $\delta_\CF$ for the space of linear continuous functionals $F:C[0;1]\to\IR$ and $[\delta_C\to\rho]$ for the subset of non-negative ones,\\
--  \ $\delta_\BV$ for the set $\BV$ of (normalized) functions of bounded variations and $\delta_{\rm V}$ for the subset of non-decreasing ones,\\
-- \ $\delta_\BM$ for the set of signed measures and $\delta_m$
for the subset of non-negative ones.

For all these representations the norm or the variation can be
computed from the names. For the representations $\delta_\CF$,
$\delta_\BV$ and $\delta_\BM$ it is included explicitly in the
names, for the other representation norms can be computed from
names: $\|F\|=F(\one)$, ${\rm Var}(g)=g(1)$, $\|\mu\|_m
=\mu([0;1])$.

Let $T_{\rm FV}$,  $T_{\rm VM}$ and $T_{\rm MF}$  be the linear
homeomorphisms from Theorem~\ref{t2} and let $T_{\rm FV}^+$,
$T_{\rm VM}^+$ and $T_{\rm MF}^+$ be their restrictions to the
spaces of non-negative or non-decreasing objects, respectively.

\begin{thm}\label{t6}$ $
\begin{enumerate}
\item \label{t6a}  The operator $T_{\rm FV}^+$ is
    $([\delta_C\to\rho],\delta_{\rm V})$-computable.

\item  \label{t6b} The operator $T_{\rm VM}^+$   is
    $(\delta_{\rm V},\delta_m)$-computable.

\item  \label{t6c}  The operator $T_{\rm MF}^+$ is
    $(\delta_m,[\delta_C\to\rho])$-computable.
\end{enumerate}
\end{thm}

\pproof
\begin{enumerate}
\item
 If $F$ is non-decreasing then $\|F\|=F( \one)$. By
Theorem~\ref{t5}, the restriction is
$([\delta_C\to\rho],\delta_{\rm V})$-computable.
\medskip

\item
Suppose, $\delta_{\rm V}(p)=g$ is non-decreasing with
dense set $A_p$ (Definition~\ref{d7}). From the classical theory
we know that for  $0\leq a<b\leq 1$ the measure $\mu:=T^+_{\rm
VM}(g)$ satisfies
\[\eqalign{\mu([0;b))&=\sup_{b'<b}g(b')\cr
           \mu((a;b))&=\sup_{a<a'<b'<b}(g(b')-g(a'))
  }
\]
  and 
\[\mu((a;1])=\sup_{a<a'}(g(1)- g(a')).\]
  Since $\lim
_{y\nearrow x}g(y)$ and $\lim _{y\searrow x}g(y)$ exist for all
$0<x<1$ it suffices to choose $a'$ and $b'$ from the dense set
$A_p$. Therefore,
\begin{eqnarray*}
\mu([0;b))&=& \sup\{g(b')\mid b'<b, \ b'\in A_p\}\,,\\
\mu((a;b))&=& \sup\{g(b')-g(a')\mid a<a'<b'<b, \ a',b'\in A_p\}\,,\\
\mu((a;1])&=& \sup \{ g(1)-g(a')\mid a<a', \ a'\in A_p\}\,.
\end{eqnarray*}
The name $p$ of $g$ contains a list of all ($(\rho,\rho)$-names
of) $(x,g(x))$ with $x\in A_p$. Since $x<y$ is r.e., for rational
numbers $a<b$ we can compute a list of all $d\in\IQ$ such that
$d<g(b')-g(a')$ for some $a',b'\in A_p$ and $a<a'<b'<b$, which is
a list of all
 $d\in\IQ$ such that $d<\mu((a;b))$.
Correspondingly, for a rational number $b>0$ we can compute a list
of all $d\in\IQ$ such that $d<\mu([0;b))$ and for a rational
number $a<1$ we can compute a list of all $d\in\IQ$ such that
$d<\mu((a;1])$. Combining these enumerations for $p$ we can
enumerate
 a list of $(d,J)\in \IQ\times{\rm Int}$ ($\rm Int$ is defined before Definition~\ref{d6}) such that $d<\mu(J)$.
Furthermore, from $p=\langle\langle p_0,q_0\rangle,\langle
p_1,q_1\rangle, \ldots\rangle$ we can compute
$\mu([0;1])=g(1)=\rho(q_1)$. Therefore, we can compute a
$\delta_m$-name of the measure $\mu$.
\medskip

\item
  This follows from Theorem~\ref{t7}(\ref{t7a}).\qed
\end{enumerate}\medskip

By the following theorems the linear homeomorphisms $T_{\rm FV}$,
$T_{\rm VM}$ and $T_{\rm MF}$  from Theorem~\ref{t2} are
computable, and Jordan decomposition on the three spaces is
computable.

\begin{thm}\label{t3}$ $
\begin{enumerate}
\item \label{t3a}  The operator $T_{\rm FV}:\Cp\to \BV$
    mapping functionals to functions of bounded variation is
    $(\delta_\CF,\delta_\BV)$-computable.
\item  \label{t3b} The operator $T_{\rm VM}:\BV\to\BM$ mapping
    functions of bounded variation to signed measures is
    $(\delta_\BV,\delta_\BM)$-computable.

\item  \label{t3c}  The operator $T_{\rm MF}:\BM\to \Cp$
    mapping signed measures to functionals is
    $(\delta_\BM,\delta_\CF)$-computable.

\end{enumerate}
\end{thm}

\begin{thm}[Computable Jordan decomposition]\label{t4}$ $
\begin{enumerate}
\item \label{t4a}  Jordan decomposition $F\mapsto (F^+,F^-)$ on $\Cp$ is \\
$(\delta_\CF,[[\delta_C\to\rho,],[\delta_C\to\rho]])$-computable. \\
Its inverse is
$([[\delta_C\to\rho,],[\delta_C\to\rho]],\delta_\CF)$-computable.
\item  \label{t4b} Jordan decomposition $g\mapsto(g^+,g^-)$ on
    $\BV$ is $(\delta_\BV,[\delta_{\rm V},\delta_{\rm
    V}])$-computable. \\Its inverse is $([\delta_{\rm
    V},\delta_{\rm V}],\delta_\BV)$-computable.

\item  \label{t4c} Jordan decomposition
    $\mu\mapsto(\mu^+,\mu^-)$ on $\BM$ is
    $(\delta_\BM,[\delta_m,\delta_m])$-computable. \\Its
    inverse is $([\delta_m,\delta_m],\delta_\BM)$-computable.
\end{enumerate}
\end{thm}

\noindent Since $f\mapsto \|f\|$ for non-negative continuous $f$ is
$([\delta_C\to\rho],\rho)$-computable in (\ref{t3a}) of the
theorem
$[\delta_C\to\rho]$ can be replaced by $\delta_{\rm CF}$. Correspondingly, in (\ref{t3b}) of the theorem $\delta_{\rm V}$ can be replaced by $\delta_{\rm BV}$ and in (\ref{t3c}) of the theorem $\delta_m$ can be replaced by $\delta_{\rm BM}$.\\

\pproof This is a merged proof of Theorems~\ref{t3} and ~\ref{t4}. Almost all statements follow easily from what has already been proved. The only non-trivial part is the proof for the Jordan decomposition $g\mapsto (g^+,g^-)$.
In the following  $(F^+,F^-)$, $(g^+,g^-)$ and $(\mu^+,\mu^-)$
will denote Jordan decompositions. By Theorem~\ref{t6} and
Corollary~\ref{c3},
\begin{eqnarray}\label{f19}(F^+,F^-)\mapsto (g^+,g^-)\mapsto (\mu^+,\mu^-)\mapsto (F^+,F^-) &\mbox{are computable}
\end{eqnarray}
w.r.t the representations $[\delta_C\to\rho]$, $\delta_{\rm V}$
and $\delta_m$.
\medskip

 {\boldmath $F\mapsto g$} (Theorem~\ref{t3}(\ref{t3a}))
  This follows immediately from Theorem~\ref{t5}.
\medskip

{\boldmath $g\mapsto (g^+,g^-)$} (first part of
Theorem~\ref{t4}(\ref{t4b}))
 Let $g\in\BV$ with Jordan decomposition $(g^+,g^-)$. From the classical theory we know $g^+=(/g/ +g)/2$ and $g^-=(/g/-g)/2$ where $/g/\in BV$ is the (non-decreasing) total variation function of $g$ (see Section~\ref{secq}
after~(\ref{f18})).

Suppose $\delta_\BV(\langle p,q\rangle)=g$.  Let $A_p$ be the
dense set from Definition~\ref{d7}. The functions $/g/$, $g^+$ and
$g^-$ are determined uniquely by their restrictions to the dense
subset $A_p\cup\{0,1\}$, hence it suffices to find $/g/(x)$,
$g^+(x)$ and $g^-(x)$ for all $x\in A_p\cup \{0,1\}$.

Call a partition $Z=(a=x_0<x_1<\ldots<x_n=b)$ of $[a;b]$ a
partition ``from $A_p$'', if $\{x_0,\ldots, x_n\}\In \{0,1\}\cup
A_p$. Suppose $x\in A_p$.

Since $g$ is left-continuous by Lemma~\ref{l8}, and $A_p$ is
dense, for every partition $Z$ of $[0;x]$ and every $\varepsilon$
there is some partition $Z'$ of $[0;x]$ from $A_p$, such that
$|S(g,Z)-S(g,Z')|<\varepsilon$. Therefore,
\[V_0^x(g)=\sup\{S(g,Z)\mid Z\mbox{ is a partition of $[0;x] $ from $A_p$}\}\]
By Definition~\ref{d7}, $p$ can be written as $p=\langle \langle
p_0,q_0\rangle,\langle p_1,q_1\rangle,\ldots\rangle$ such that
$\rho(p_0)=\rho(q_0)=0$, $\rho(p_1)=1$ and $g\circ
\rho(p_k)=\rho(q_k)$ for all $k\in\IN$.

Let $x_k:= \rho(p_k)$ and $y_k:=\rho(q_k)=g(x_k)$. We want to
compute  a sequence $t:=\langle \langle p_0,r_0\rangle,\langle
p_1,r_1\rangle,\ldots\rangle$ such that
$\rho(r_k)=/g/(x_k)=V_0^{x_k}(g)$. Since $/g/$ is continuous in
$x$ if $g$ is continuous in $x$, then $\delta_{\rm V}(t)=/g/$.

Since $/g/(0)=0$ we can choose $r_0:= q_0$. Since $/g/(1)={\rm
Var}(g)=\rho(q)$ (remember that $\delta_\BV(\langle
p,q\rangle)=g$) we can choose $r_1:= q$.

For $k\geq 2$ let $\Pi(k)$ be the set of all sequences
$\sigma=(i_0,i_1,\ldots, i_m)$ such that $i_0=0$ $i_m=k$ and
$x_{i_0}<x_{i_1}<\ldots <x_{i_m}$. For $\sigma=(i_0,i_1,\ldots,
i_m)$ let $P_\sigma$ be the partition $(x_{i_0},x_{i_1},\ldots
,x_{i_m})$ of $[0,x_k]$ from $A_p$. Then
$V_0^{x_k}(g)=\sup_{\sigma\in \Pi(k)} S(g,P_\sigma)$.

Since the relation $x<y$ for real numbers is
$(\rho,\rho)$-enumerable, from $p$ and $k$ the set $\Pi(k)$ can be
enumerated, $\Pi(k)=(\sigma_0,\sigma_1,\ldots)$. Since
$S(g,P_\sigma)$ can be computed from $p$ and $\sigma$ (\ref{f17}),
a $\rho_<$-name of $V_0^{x_k}(g)=\sup_k S(g,P_{\sigma_k})$ can be
computed from $p$ and $k$. Correspondingly, a $\rho_<$-name of
$V_{x_k}^1(g)=\sup\{S(g,Z)\mid Z\mbox{ is a partition of $[x_k;1]
$ from $A_p$}\}$ can be computed from $p$ and $k$.

Since $V_0^x(g)+V_x^1(g)=V_0^1(g)={\rm Var}(g)$ and  $\rho$-name
of ${\rm Var}(g)$ is given as an input, from $p$,  $q$ with
${\rho(q)=\rm Var}(g)$
 and $k$ a $\rho$-name $r_k$ of $/g/(x_k)$ can be computed. Therefore, some computable function $G\pf \om\to\om$ maps every $\delta_\BV$-name $\langle p,q\rangle$
  of $g$ where $\delta_{\rm V}(p)=g$ and  $p=\langle \langle p_0,q_0\rangle,\langle p_1,q_1\rangle,\ldots\rangle$
 to some $\delta_{\rm V}$-name $t=\langle \langle p_0,r_0\rangle,\langle p_1,r_1\rangle,\ldots\rangle$ of $/g/$.

On these names, $g^+=(/g/ +g)/2$ and $g^-= ( /g/-g)/2$ can be
computed: there are computable functions $s,d$ on $\om$ such that
$(\rho(p)+\rho(q))/2=\rho\circ s(p,q)$ and
$(\rho(p)-\rho(q))/2=\rho\circ d(p,q)$. Then $t^+:=\langle \langle
p_0,s(q_0,r_0)\rangle,\langle p_1,s(q_1,r_1)\rangle,\ldots
\rangle$ is a $\delta_{\rm V}$-name of $g^+$ and $t^-:=\langle
\langle p_0,d(q_0,r_0)\rangle,\langle p_1,d(q_1,r_1)\rangle,\ldots
\rangle$ is a $\delta_{\rm V}$-name of $g^-$. In summary, $t^+$
and $t^-$, hence $\langle t^+,t^-\rangle$ can be computed from
$\langle p,q\rangle$. Therefore, Jordan decomposition
$g\mapsto(g^+,g^-)$ on $\BV$ is $(\delta_\BV,[\delta_{\rm
V},\delta_{\rm V}])$-computable.
\medskip

{\boldmath $(\mu^+,\mu^-)\mapsto\mu$} (second part of
Theorem~\ref{t4}(\ref{t4c})) By Theorem~\ref{t14}, from
$\delta_m$-names of a Jordan decomposition $(\mu^+,\mu^-)$ we can
compute a $\delta_\BM$-name of $\mu$.
\medskip

{\boldmath $g\mapsto \mu$} (Theorem~\ref{t3}(\ref{t3b})) Compute
as follows: $g\mapsto (g^+,g^-)\mapsto (\mu^+,\mu^-)\mapsto\mu$.
\medskip

{\boldmath $\mu\mapsto F$} (Theorem~\ref{t3}(\ref{t3c})) Suppose
$\delta_\BM\langle p,q,r\rangle =\mu$, hence $\mu=\mu^+-\mu^-$
where $\mu^+=\delta_m(p)$ and $\mu^-=\delta_m(q)$ and
$\|\mu\|_m=\rho(r)$. By Theorem~\ref{t6}(\ref{t6c}) we can compute
$[\delta_C\to\rho]$-names of functionals $G^+:=T_{\rm MF}(\mu^+)$
and $G^-:=T_{\rm MF}(\mu^-)$ such that $F:= T_{\rm
MF}(\mu)=G^+-G^-$. By a standard argument we can compute a
$[\delta_C\to\rho]$-name of $F$. Since $\|F\|=\|\mu\|_m$ by
Theorem~\ref{t2}, we can compute a $\delta_\CF$-name of $F$.
\medskip

{\boldmath $(g^+,g^-)\mapsto g$} (second part of
Theorem~\ref{t4}(\ref{t4b})) Compute as follows: $(g^+,g^-)\mapsto
(\mu^+,\mu^-)\mapsto\mu\mapsto F\mapsto g$.
\medskip

{\boldmath $\mu\mapsto (\mu^+,\mu^-)$} (first part of
Theorem~\ref{t4}(\ref{t4c})) Compute as follows: $\mu\mapsto
F\mapsto g\mapsto (g^+,g^-)\mapsto(\mu^+,\mu^-)$.
\medskip

{\boldmath $F\mapsto (F^+,F^-)$} (first part of
Theorem~\ref{t4}(\ref{t4a})) Compute as follows: $F\mapsto
g\mapsto (g^+,g^-)\mapsto (F^+,F^-)$.
\medskip

{\boldmath $ (F^+,F^-)\mapsto F$} (second part of
Theorem~\ref{t4}(\ref{t4a})) Compute as follows: $(F^+,F^-)\mapsto
(g^+,g^-)\mapsto g\mapsto\mu\mapsto F$.
\qed

\begin{cor}\label{c5}$ $
\begin{enumerate}
\item \label{c5c}The inverses $(T_{\rm FV}^+)^{-1}$, $(T_{\rm
    VM}^+)^{-1}$  and $(T_{\rm MF}^+)^{-1}$ are computable.

\item \label{c5a}The inverses $T_{\rm FV}^{-1}$, $T_{\rm
    VM}^{-1}$  and $T_{\rm MF}^{-1}$ are computable.
\item \label{c5b}$\delta_\CF\equiv T_{\rm FV}^{-1}\circ
    \delta_\BV\equiv T_{\rm MF}\circ\delta_\BM$ (accordingly
    for $\delta_\BV$ and $\delta_\BM$).
\end{enumerate}
\end{cor}

\pproof\hfill

\begin{enumerate}
\item For all $F\in C'[0;1]$, $T_{\rm MF}^+\circ T_{\rm
VM}^+\circ T_{\rm FV}^+(F)=F$, hence $T_{\rm MF}^+\circ T_{\rm
VM}^+ = (T_{\rm FV}^+)^{-1}$ which is computable by
Theorem~\ref{t6}. The other statements are proved accordingly.

\item As above, but with Theorem~\ref{t3}.

\item Straightforward by \ref{c5a}. and Theorem~\ref{t3}.\qed
\end{enumerate}

We introduce further representations of our spaces by differences
of functions:
\smallskip

\begin{itemize}
\item $\gamma_{\rm F}\langle p,q,r\rangle =F$ iff $F=[\delta_C\to\rho](p)-[\delta_C\to\rho](q)$ and $\|F\|=\rho(r)$,\\
\item $\gamma_{\rm V}\langle p,q,r\rangle =g$ iff $g=\delta_{\rm V}(p)- \delta_{\rm V}(q)$ and ${\rm Var}(g)=\rho(r)$,\\
\item $\gamma_{\rm FJ}\langle p,q\rangle =F$ iff  ($[\delta_C\to\rho](p),  [\delta_C\to\rho](q))$ is the Jordan decomposition of~$F$,\\
\item$\gamma_{\rm VJ}\langle p,q\rangle =g$ iff $(\delta_{\rm V}(p),\  \delta_{\rm V}(q))$ is the Jordan decomposition of $g$,\\
\item$\gamma_{\rm MJ}\langle p,q\rangle =\mu$ iff
    $(\delta_m(p),\delta_m(q))$ is the Jordan decomposition of
    $\mu$.
\end{itemize}
\medskip

\begin{cor}\label{c4}\ 
$\delta_\CF\equiv\gamma_{\rm F}\equiv\gamma_{\rm FJ}$, \ \
$\delta_\BV\equiv\gamma_{\rm V}\equiv\gamma_{\rm VJ}$, \ \
$\delta_\BM\equiv \gamma_{\rm MJ}$.
\end{cor}

\pproof Straightforward by Theorems~\ref{t3} and~\ref{t4}.
\qed

Notice that for each of these representations a name of a
functional $F$ contains a name of $\|F\|$ or allows to compute it
easily. On the Banach spaces $(\Cp,\|\,.\,\|)$, $(\BV,{\rm Var})$
and $(\BM,\|\,.\,\|_m)$ with representations $\delta_\CF$,
$\delta_\BV$ and $\delta_\BM$, respectively, addition is not
computable since the norm of the sum cannot be computed. Adding
the norm in a representation of the dual space is discussed in
\cite{Bra01b,Bra05d,BS05}. But for non-negative functionals $F$:

\begin{cor} \label{c1} The sum
\begin{enumerate}
 \item\label{c1a} of  non-negative functionals from
     $\,C'[0;1]$ is computable w.r.t. $[\delta_C\to\rho]$,
\item\label{c1b} of non-decreasing functions from $\BV$ is
    computable w.r.t $\delta_{\rm V}$,
\item\label{c1c} of non-negative bounded measures from $\BM$
    is computable w.r.t $\delta_m$.

\end{enumerate}
 \end{cor}

\pproof\hfill

\begin{enumerate}
\item Straightforward \cite[Theorem~6.2.1]{Wei00}.

\item This follows from  \cite[Theorem~3.1]{Wei99a}.

\item Since for non-decreasing $g_1,g_2$, \ $g_1+g_2=(T_{\rm
VM}^+)^{-1}(T_{\rm VM}^+(g_1)+T_{\rm VM}^+(g_2))$, by \ref{c1c}.
of this corollary, Theorem~\ref{t6} and Corollary~\ref{c5} the sum
on non-decreasing functions is computable w.r.t. $\delta_{\rm V}$.\qed
\end{enumerate}

\noindent For functions of bounded variation there is no simple proof since
for $g_1=\delta_{\rm V}(p_1)$ and $g_2=\delta_{\rm V}(g_2)$ in
general $A_{p_1}\neq A_{p_2}$ (see Definition~\ref{d7}).

\section{Acknowledgement}
The authors thank the unknown referees for their careful work.

\bibliographystyle{plain}

\end{document}